\documentclass[showpacs,amsmath,amssymb,twocolumn,aps,
        superscriptaddress]{revtex4}

 \usepackage{graphicx}
\usepackage{epsfig}
 \usepackage{pstricks}
 \usepackage{pst-plot}
  \usepackage{psfrag}
 \usepackage{dcolumn}
 \usepackage{bm}

\usepackage{color}

\def\mn#1{}

\def\lfrac#1#2{#1/#2}

\begin{document} 
\title{Strong Enhancement of High Voltage  
Electronic Transport in Chiral Electrical Nanotube 
Superlattices}

\author{J\"urgen Dietel}
\affiliation{Institut f\"ur Theoretische Physik,
Freie Universit\"at Berlin, Arnimallee 14, D-14195 Berlin, Germany}
\author{Hagen Kleinert}
\affiliation{Institut f\"ur Theoretische Physik,
Freie Universit\"at Berlin, Arnimallee 14, D-14195 Berlin, Germany}
\affiliation{ICRANeT, Piazzale della Repubblica 1, 10 -65122, Pescara, Italy}
\date{Received \today}

\begin{abstract}
We consider metallic carbon nanotubes with an 
overlying unidirectional electrical chiral 
(wavevector out of the radial direction, where the axial direction 
is included)
superlattice potential. We show that for  
superlattices with a wavevector close to the axial direction, 
the electron velocity 
assumes the same value as for nanotubes without superlattice. 
Due to an increased number of phonons with 
different momenta  but lower electron-phonon scattering 
probabilities, we obtain a large enhancement of  
the high-voltage conductance and current sustainability in comparison with 
the nanotube without superlattice.        
   
\end{abstract}

\pacs{63.22.Gh, 73.21.Cd, 73.63.Fg, 73.50.Fq}

\maketitle

Depending on their chirality, carbon nanotubes (NT) behave 
either like a semi-conductor
or a metal. In the first case, they offer
 interesting alternative 
for building logical circuits. In the second case, they can be used 
as nanometer-sized metallic wires in logical circuits. This is
particularly useful since they can 
sustain very high currents before breaking.
At low voltages ($U \lesssim 0.17 V $) 
the effective electron scattering length at room temperature 
in metallic NTs 
is mainly governed by 
acoustical phonon and impurity scattering 
with a value of a few hundred nanometers \cite{Purewal1}.
At higher voltages, scattering with hot optical phonons
 created by electron-phonon scattering \mn{OK?} 
becomes relevant. This leads
to a significant reduction of the electron's
mean free path 
down to roughly $ l_{\rm sc} \approx 10 $ nm 
\cite{Yao1, Park1, Javey1, Lazzeri1, Dietel1}, resulting in
a large increase in the absolute and 
differential resistance. Due to the 
large number of optical phonons, phonon-phonon scattering with acoustic 
phonons produces heat in the NT that
ultimately causes the electrical breakdown \cite{Collins1,Huang1}. 

In Ref.~\onlinecite{Vandecasteele1} it was argued that 
the performance of a metallic NT, i.e., 
its absolute and differential conductance,  
can be enhanced considerably by isotopical disorder enrichment. 
This causes
additional relaxation 
paths for optical phonons by disorder scattering.   
The purpose of the present letter is 
 to propose a different 
 mechanism \mn{OK?} to enhance the electronic transport. 
We show that by applying an unidirectional electrical 
superlattice (SL) (cf. Fig.~1)
with wavevector close to the axial direction  
of the NT, 
we can enhance the (differential) conductance 
 considerably, especially in the large 
voltage regime. Such a potential could be for example  
produced by  adatom deposition via electron beams 
directed on the NT \cite{Meyer1} or (at least approximately) 
by twisted periodical patterned top and bottom  gate electrodes
or the coupling of the NT to surface acoustic waves \cite{Talyanskii1}.   
Since the (average) phonon number is  
proportional to the inverse 
electron-phonon scattering time $ 1/ \tau_{\rm ep} $,
and the inverse electron mean-free path $ 1/ l_{\rm sc} $ 
is proportional to the phonon number 
times $ 1/ \tau_{\rm ep} $ in the hot phonon regime, we obtain   
a quadratic dependence of the electron mean-free path on the  
scattering time $l_{\rm sc}  \sim  \tau^2_{\rm ep} $. 
Below it will be shown
 that an application of an electrical chiral potential 
causes
a large number of different 
phonons to take part in the 
electron-phonon scattering process with increased electron-phonon 
scattering times, 
so that $ 1/ \tau_{\rm ep} \sim \sum_{i} 1/ \tau^i_{\rm ep} $. 
This is what causes the strong decrease of the (differential) 
resistance  that scales with $ 1/l_{\rm sc}  \sim 
\sum_i 1/ (\tau^i_{\rm ep})^2 \ll 1/ \tau^2_{\rm ep}$ by using 
Matthiessen's rule and reduces the phonon temperature\mn{OK?}.    

It was shown recently that new Dirac points 
in the energy spectrum can be opened by imposing an 
SL on the graphene lattice \cite{Park2, 
Brey1, Arovas1, Barbier1, Wang1}. 
This is also seen in NTs for potentials with wavevector 
in the radial direction. 
We will show that they vanish for general chiral potentials.   

The Hamiltonian near the Dirac point $ {\bf K} $ 
for a NT with axis in $ y' $ direction subjected to an SL potential 
reads
\begin{equation} 
H_K = \left(\begin{array} {c c}
V(x'+ t_\gamma y') 
 &-i{\hbar v_F} \left(\partial_{x'}-i \partial_{y'}\right)   \\
-i{\hbar v_F} \left(\partial_{x'}+i \partial_{y'}\right) &  
V(x'+ t_\gamma y')  
\end{array} \right) \,, \label{20}
\end{equation} 
where $ v_F $ is the Fermi velocity.
For a NT with circumference D, the SL potential $ V $ is periodic in the 
radial (axial) direction with periodicity $ d $ ($d/t_\gamma$), i.e.,
  $ V(x'+d) = V(x') $ and one has $ D/d \in \mathbb{N} $.
In the following we  solve  
the eigenvalue equation $ H_K {\bf u}({\bf r}')=\epsilon \,  
{\bf u}({\bf r}') $.  
We use the abbreviation $ t_\gamma= \tan( \gamma) $ where 
$ \gamma $ is the chiral angle of the SL potential 
$ V $. The metallic NT boundary conditions are given by 
$ {\bf u}(x'+D,y')={\bf u}(x',y')$ \cite{Ajiki1}. 
 
To solve the eigenvalue equation we follow first a transfer matrix 
method similar to Ref.~\onlinecite{Arovas1}. 
By using the coordinates $ x=x'+t_\gamma y' $ and $ y=y' $,
the solution of the Schr\"odinger equation has the Bloch form\mn{OK?}
$ {\bf u}({\bf r})= e^{i q y}(u_1(x),u_2(x))^T $ with $  
(u_1(x), u_2(x))^T = \Lambda(x) (u_1(0), u_2 (0))^T $,     
where 
\begin{align} 
& \Lambda(x)=e^{-i \tilde{q}x} {\cal P} 
\exp\left[\int_0^{x} dx'M_{V(x')}\right]\,,   \label{30}  \\
& M_{V(x)} = \left( 
\begin{array}{c c}
 \lfrac{q}{T_\gamma^2} & i \lfrac{\kappa(x)}{(1+i t_\gamma)}  \\
 i \lfrac{\kappa(x)}{(1-i t_\gamma)} & -\lfrac{q}{T_\gamma^2} \,,
\end{array} 
\right)  \label{60}
\end{align}
and $ \tilde{q}  =  q \lfrac{t_\gamma}{T_\gamma^2} $, 
$ \kappa(x)= [\epsilon -V(x)]/\hbar v_F $, $
T_\gamma=\sqrt{1+t_\gamma^2}$.
The operator  
 $ {\cal P} $ indicates path ordering and  places
all larger 
values of $ x $ to the left\mn{OK?}.
The Bloch condition reads 
$ (u_1(d),u_2(d))^T= e^{i \eta} (u_1(0),u_2(0))^T $ with 
 $ \rm{det}[e^{i \eta}-\Lambda(d)]=0 $ when $ d $ is the SL wavelength.

Consider first a carbon NT in a chiral periodical 
lattice of two piecewise constant potentials of the form 
\begin{equation} 
V(x)= \left\{ \begin{array}{ c c c}
V_1 & \mbox{if} & 0 \le x < d_1 \\
V_2 & \mbox{if} & d_1 \le x < d_1+d_2  \,.\\
\end{array} \right.
\label{80} 
\end{equation}  
Then we obtain $ \Lambda(d)= \Lambda_1 \Lambda_2 $, where $d=d_1+d_2 $.    
\begin{align} 
& \Lambda_i=e^{-i\tilde{q}d_i} 
\left\{\cos[\alpha_\epsilon(d_i)] +\frac{\sin[\alpha_\epsilon(d_i)]}
{\alpha_\epsilon(d_i)} M_{V_i} d_i \right\}  \,, 
\label{90}  \\ 
& 
\alpha_\epsilon(d_i)=d_i \sqrt{\kappa_i^2/T_\gamma^2-
      q^2/T_\gamma^4} \,. \label{100}
\end{align} 
$ \kappa_i $ is given by $ \kappa(x) $,  with $ V(x)=V_i $. Since 
$ \rm{det}[e^{i\tilde{q}x} \Lambda(x)]=1 $ we have for the eigenvalues of the 
matrix $ \Lambda(d)$,  $ \xi=e^{-i\tilde{q} d}(1/2)(T \pm \sqrt{T^2-4}) $, with 
$ T  =\rm{Tr}[e^{ i\tilde{q}d} \Lambda(d) ] $ and 
\begin{align} 
  T &= 2 \cos[\alpha(d_1)] \cos[\alpha(d_2)]  \label{110}  \\
&  +  
 2 \frac{\sin[\alpha(d_1)] \sin[\alpha(d_2)]}{\alpha(d_1) \alpha(d_2)} 
\left( \frac{q^2}{T^4_\gamma}
-\frac{\kappa_1 \kappa_2}{T^2_\gamma} \right) d_1 d_2 \,.
 \nonumber  
\end{align}        
By taking into account that $ T $ is real, we obtain from 
(\ref{110}) the dispersion relation 
\begin{equation} 
2 \cos( \tilde{q}  d+ \eta )= T  \,.       \label{120}  
\end{equation} 
 
In the following, we restrict ourselves 
to mirror symmetric potentials $ V_1=-V_2=V $, with $ d_1=d_2=d/2 $ leading to 
the best current-voltage results over all two-step potentials.  
This leads to an energy spectrum that possesses 
a mirror symmetry at $ \epsilon=0 $
as a function 
of the quasi-momentum $ q $ in y-direction.
For $ \epsilon = 0 $, we obtain from Eq.~(\ref{120}) that  
$ T= 2 + q^2 d^2 \sin^2[\alpha_0(d/2)]/\alpha_0^2(d/2) T_\gamma^4 $.
This leads with (\ref{120}) to the
existence  of new Dirac points \cite{Arovas1}
at zero chirality  
$ t_\gamma = 0 $. \mn{check}
The number of these points is given by 
$ [V d/ 2 \pi \hbar v_F T_\gamma] $ 
where $[x] $ is the largest integer number smaller than $ x$. For 
$ t_\gamma \not =0 $ an energy gap is opened and the  
Dirac points disappear (see left panel in Fig.~1).\mn{Is this what you want
  to say? What is a vanishing point?} 
\begin{figure}
\begin{center}
\includegraphics[clip,height=0.8cm,width=7.0cm]{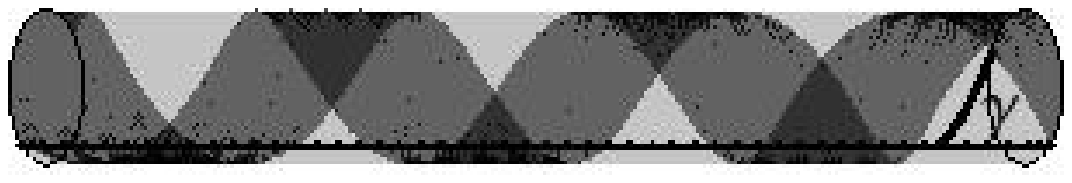}\\[0.08cm]
\includegraphics[clip,height=6.0cm,width=8.5cm]{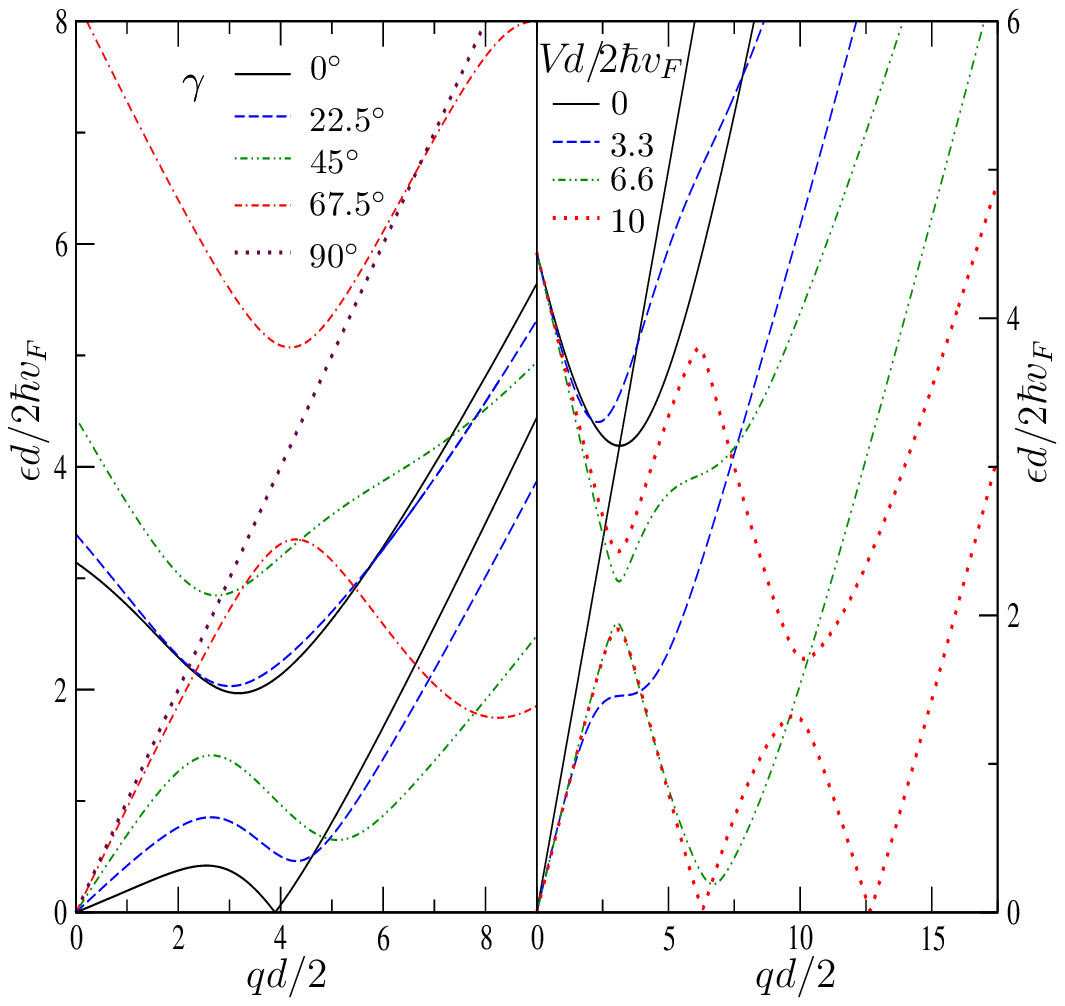}
\vspace*{-0.2cm}
\caption{(Color online) 
Upper panel shows a NT with an overlying chiral electrical superlattice 
potential.  
Lower panel shows the two lowest energy bands ($d=D$) 
by solving Eq.~(\ref{120}) 
for $ \eta=0 $ as a function of the rescaled axial quasi-momentum 
$ q d/2 $ for various chiral angles $ \gamma $ 
with $ Vd/ 2 \hbar v_F =5 $ (left panel)  
and various chiral potentials $ Vd/2 \hbar v_F $ 
with $ t_\gamma=1 $ (right panel). Note that the 
dotted curve in the right panel crosses the x-axis exactly only at $ q=0 $.   
}
\vspace{-0.3cm}
\end{center}
\end{figure}

Next we calculate the energy values of the bands at zero momentum $ q=0 $. 
Eqs.~(\ref{110}) and (\ref{120}) delivers for these energy values
$ \epsilon^\eta(0)= (\pm \eta+2 \pi n) T_\gamma \hbar v_F/d $
where $ n $ determines the energy bands for fixed quasimomentum $ \eta $. 
Thus the energy bands are far more separated in energy 
space for $ t_\gamma \gg 1 $  than for the 
system without chiral potential, i.e. $ \epsilon^\eta(0) $ for 
$ t_\gamma =0  $. 
   
In order to see how the lowest energy band scales with 
$V $ and $ t_\gamma $, 
we calculate from (\ref{110}) the energy dispersion 
of the lowest band for metallic NTs, i.e. $ \eta=0$, in the regime  
$  |\epsilon_s| \ll \hbar v_F T_\gamma/d  $, $V $ and   
$ q^2 \ll T^2_\gamma  
(V(x)/\hbar v_F)^2 $, to be called ${\cal  R} $.   
We then obtain 
 \begin{align} 
&  \epsilon_s  = \! \!  s \hbar v_F  
 \sqrt{ 
  \frac{|q \Gamma|^2}{T_\gamma^2}
   + 4 
 \frac{T_\gamma^2}{d^2}
      \sin^2  \left(\frac{\tilde{q} d}{2} \right)
 }   \,  ,         \label{130} \\
& 
 \Gamma= \frac{1}{d} \int_0^d  dx 
\exp\left[i \, 2 \int^x_0 dx' {\rm sgn}[V(x')] \alpha_0(x')/x'    \right]  \label{135}
\end{align}  
where $ s=\pm 1$ and  
$ {\rm sgn}[x] $ is the sign of $ x $. Note that 
$ \Gamma=\sin[\alpha_0(d/2)]e^{i \alpha_0(d/2)}/\alpha_0(d/2) $ for the symmetric 
two-step potential (\ref{80}). 

In Fig.~1, we plot the two lowest energy bands ($d=D$)
for $ \eta=0 $, by solving 
Eq.~(\ref{120}) numerically.
Eq.~(\ref{130}) leads to the electron velocity 
$ v_{y'}(q)=\partial \epsilon_s/ \partial (\hbar q) $ along the NT axis
(the $q$ dependency of $ \tilde{q} $ has to be considered in the derivate). 
We restrict our discussion to momentum region near the Dirac point, i.e., 
$ \tilde{q}d/2 \ll 1 $.   
It is  smaller for larger potentials $ V $ being maximal 
at $ \alpha_0 \to 0 $ with 
value $ v_{y'}(q)  \le  v_F $.
We point out that in general for $ |t_\gamma| \gtrsim 1  $ we have 
$ v_{y'}(q) \approx  
v_F $ in $ {\cal R} $ 
irrespective of the potential strength $ V $. Since electron-phonon 
scattering times are proportional to $ v_{y'}(q) $ we restrict our 
transport calculations below to NTs with 
$ |t_\gamma| \gtrsim 1  $ leading to the largest conductivities.  
For the group velocity of the electrons in radial direction we have 
$ v_{x'}(q)= \partial \epsilon_s/ \partial (\hbar \tilde{q})
- t_\gamma v_{y'}(q)   $.  
For $ t_\gamma \gtrsim 1 $ we obtain  
$ v_{x'}(q) \approx  v_F (1-|\Gamma|^2)/T_\gamma $
in $ {\cal R} $ leading to the collimination of the electron beam 
\cite{Park3}, i.e., $|v_{y'}(q)|> |v_{x'}(q)| $. 
This expression is even valid where 
$ t^2 \gg  |\Gamma|^2  $.    
On the other hand for $ t^2 \ll |\Gamma|^2  $ we
obtain $ v_{x'} \approx t_\gamma v_F (1- |\Gamma|^2)/\Gamma$ leading to 
the vanishing of collimination at chiral angles $ |\Gamma|^2/(1- |\Gamma|^2) 
\lesssim |t_\gamma| \lesssim (1-|\Gamma|^2) 
 $. For even smaller $ t_\gamma $ we 
obtain collimination again. 
         
In Fig.~1 we see how 
the energy bands oscillate, in accordance 
with  Eq.~(\ref{130}) for $ q  \le T_\gamma V/\hbar v_F $, 
thus forming band SL valleys.
The central valley possesses a true Dirac point at $ q=0 $. The SL 
side valleys  have then a minimum at 
$ \sin^2  (\tilde{q} d/2) =0 $.  
Within the SL valleys, electrons travel either to the left (right) for 
$ \partial \epsilon_s/\partial q < 0 $ 
($ \partial \epsilon_s/\partial q > 0 $).  The number of 
side valleys can be read of from (\ref{130}) as $ 2 m_1 $ with  $ m_1= 
[t_\gamma V d/ T_\gamma  \hbar v_F 2  \pi] $.

For SLs with 
wavevector in exact axial direction we can read off the physics  
from the chiral case 
by choosing   $ \tilde{q} \rightarrow  k $, $ t_\gamma=\eta=0 $  
where $ d $ is now the wavelength of the SL potential in 
axial direction with quasimomentum $ k $.  
The wavevector $q $ in the circumferential direction is quantized by 
$ q= 2 \pi n/ D $ due to the periodic boundary conditions of the 
wavefunctions. We point out that also the wavefunctions 
Eqs. (\ref{150}), (\ref{185}), and the 
considerations below  Eqs.~(\ref{150}), (\ref{240})  
are still valid with the additional replacements 
$ k_x \rightarrow k_y$ and $ x \rightarrow y $.      
Eq.~(\ref{130}) delivers that the energy bands are separated by 
$ \hbar v_F q \Gamma $ which means that the energy spacing between the 
energy bands goes to zero for infinite potential strength.        
From Eq.~(\ref{130}) we have $ v \rightarrow  v_F $ for 
$ V d /\hbar v_F \to \infty $ where for the lowest band, i.e.  $ q= 0 $, 
$ v \rightarrow  v_F $ even for finite potentials $ V d /\hbar v_F $.

Next, we determine the eigenvectors 
$ {\bf v}^{\eta} $ of the matrix $ \Lambda(d) $. 
These are given by $   {\bf v}^{ \eta}= 
\frac{1}{N_\Lambda} ([a + i \sin(\eta +\tilde{q} d)] /b, 1)^T $ 
with 
\begin{align}  
\!\!\! a&= \left[q d_2 \cos(\alpha_1) \frac{\sin(\alpha_2)}{\alpha_2} 
 +q d_1 \cos(\alpha_2) \frac{\sin(\alpha_1)}{\alpha_1} \right]  
\! \frac{1}{T_\gamma^2} ,  
\nonumber  \\  
\!b&=\!\left[\kappa_2  d_2 \cos(\alpha_1) \frac{\sin(\alpha_2)}{\alpha_2} 
 \!+\!\kappa_1 d_1 \cos(\alpha_2) \frac{\sin(\alpha_1)}{\alpha_1} \right]  
\!\!   \frac{i}{1- i t_\gamma}    \nonumber \\
 &-i \, \frac{\sin(\alpha_1)\sin(\alpha_2)}{\alpha_1 \alpha_2}
 \frac{d_1 d_2 \, q \, (\kappa_1-\kappa_2)}{T_\gamma^2(1- i t_\gamma)} 
\label{150}    
\end{align}      
and $ N_\Lambda $ is a normalization factor.
The eigenfunction $ {\bf u}^{\eta}(x,q) $ 
of the full Hamiltonian $ H_K $ is then given by  
$ {\bf u}^{\eta}(x,q)= (\cos[\alpha_0(x)] I + 
\{\sin[\alpha_0(x)]/\alpha_0(x)\} M_V)
{\bf v}^{\eta} $ for $ x < d/2 $. 
Eq.~(\ref{150}) leads to  
$ \langle {\bf u}^{\eta}(x,q)| e^{i k_x x} I 
 | {\bf u}^{-\eta}(x,-q) \rangle 
=0 $        
\mn{same question!!}
where we used  the abbreviation 
$ \langle {\bf u}^{\eta}(x,q)| e^{i k_x x} \sigma 
| {\bf u}^{-\eta}(x,-q) 
\rangle \equiv \int_0^d d x \, \langle {\bf u}^{\eta}(x,q)| e^{i k x} 
\sigma  | {\bf u}^{-\eta}(x,-q)\rangle $ with  
$ \sigma \in \{ \sigma_x, \sigma_y, I\} $. Here 
$ I $ is the identity matrix and 
$ \sigma_x $, $ \sigma_y $ the spin matrices.   
The wavevector of the phonons or impurities is denoted 
by $k_x $. This is a generalization of 
the results that inner-valley backward 
impurity scattering in Refs.~\onlinecite{Ando1, McEuen1}
and deformation potential phonon scattering in Ref.~\onlinecite{Suzuura1} 
does not exist 
in the lowest band in metallic NTs in contrast to semiconducting ones.

We then  obtain from (\ref{150}) for the lowest band eigenfunctions  
$ {\bf u}_{s}(x,q)$ for $ \eta=0 $  in the regime ${\cal R} $   
corresponding to the eigenvalues (\ref{130}) 
\begin{align}
&  {\bf u}_{s}(x,q) = \frac{1}{N_u} e^{-i\tilde{q}x} \bigg[
-i  \binom{\sqrt{1- i t_\gamma}}
 {\sqrt{1 +i t_\gamma}} 
\frac{T_\gamma}{\Gamma^*}  \label{185}   \\
& \times \! \!  \left( \!  \frac{T_\gamma}{q d} 
\sin(\tilde{q} d)   
+\frac{\epsilon_s}{\hbar v_F q} \right) \! \phi^*(x) 
 + \! \binom{-\sqrt{1- i t_\gamma}}{ {1 +i t_\gamma}}   \phi(x)
\bigg]\,,
\nonumber 
\end{align} 
where $ N_u $ in (\ref{185}) denotes  a normalization factor.
The phase factor $ \phi(x)$ is given by  $ \phi(x)= \exp\left[i \int_0^x dx' 
{\rm sgn}\, [V(x')] \alpha_0(x')/x'  \right] $.  
We point out that the lowest band eigenvalues 
(\ref{130}) and eigenfunctions  (\ref{185}) are more generally
valid for chiral potentials $ V(x) = V(x+D)$, where we  
have to assume that 
 $ \int_0^D dx' {\rm sgn}\, [V(x')] \alpha_0(x')/x'  =0 $. 
In order to derive  the eigenfunctions (\ref{185})  
we first formulate the eigenvalue problem corresponding to (\ref{20}) 
in the basis 
$   ( \sqrt{1- i t_\gamma},\sqrt{1+ i t_\gamma} )^T 
e^{- i \tilde{q} x} \phi^*(x) e^{iq y} $ and 
$  (- \sqrt{1- i t_\gamma},\sqrt{1+ i t_\gamma} )^T 
e^{ -i \tilde{q} x} \phi(x) e^{iq y} $. The resulting Hamiltonian is evaluated 
perturbatively  in lowest order in 
$ (\hbar v_F/d)T_\gamma  \sin(\tilde{q} d) \sigma_z    $ 
and $ (\hbar v_F q/T_\gamma) (\mbox{Re}[\phi^2(x)] \sigma_y +
\mbox{Im}[\phi^2(x)] \sigma_x) $ resulting 
in (\ref{130}) and (\ref{185}) when the 
$ \sim q^2 $ terms in $\alpha_0 $ are absent.  
The full expressions 
(\ref{130}) and (\ref{185}) 
valid in the regime $ {\cal R} $ can then be read off by comparing 
the first order expressions with the formal solution of (\ref{120}) for 
general chiral  potentials $ V(x) $ leading effectively to the 
$ \sim q^2 $ correction factor in $ \alpha_0 $ (\ref{100}).                 

By using (\ref{185}) we are now able 
to calculate the backward squared transition matrix elements 
in the regime ${\cal R} $ being inverse proportional to the inverse
electron phonon scattering time.  
Here we restrict us to SL inner and inter-valley backward scattering, i.e.
 $ \partial_{q}\sin(\tilde{q}d/2)^2 \gtrless 0 $ and 
$ \partial_{q'}\sin(\tilde{q}'d/2)^2 \lessgtr 0 $, 
at large chiral angle  $ |t_\gamma| \gtrsim  1$,  
for $ (qd)^2 \Gamma^2 \ll 
4 T_\gamma^4 \sin(\tilde{q}d/2)^2$ and 
$ (q'd)^2 \Gamma^2 \ll  4 T_\gamma^4 \sin(\tilde{q}'d/2)^2$ which 
is the relevant regime for transport at high applied bias voltages.
We obtain 
\begin{align} 
&  |\langle {\bf u}_s(x,q)|  \sigma e^{i k_x x}  | {\bf u}_{s'}(x,q')\rangle|^2 
 \approx   
(A^0_{\sigma})^2 \delta_{\tilde{k}_x d,0} \label{240} \\ 
& \qquad +  
 \frac{1}{2 m_2} (A^1_{\sigma})^2 \sum\limits^{m_2}_{j=1}
\sum_{\pm} \delta_{\tilde{k}_x d, \pm 2 \pi \left[
 2 V_j d /2 \pi \hbar v_F T_\gamma\right]} \,,
\nonumber 
\end{align} 
where $ (A^0_{\sigma})^2 \approx 0 $ 
and $ (A^1_{\sigma_x})^2 \approx 1 $, $ (A^1_{\sigma_y})^2 \approx (A^1_{I})^2
\approx 0 $. 
Here we introduced the abbreviation  
$ \tilde{k}_x=k_x- \tilde{q}'+\tilde{q} $. 
Note that we used in (\ref{240}) 
the approximation that $ \alpha_0(d_i) \approx 
d_i |V_i|/\hbar v_F T_\gamma $ valid for the majority of SL valleys. 
  
For the symmetric two-step potential (\ref{80}) 
we have $m_2=1 $ and $ V_1=V $ in (\ref{240}). 
We generalized in (\ref{240}) 
our results to a chain of symmetric two-step potentials 
with $ d_1=d_2 =d/2 m_2 $ of potential heights $ V_i $ where we assume
the potential heights are separated considerably   
$ [2 V_i d /2 \pi \hbar v_F T_\gamma] \not= 
 [ 2 V_j d /2 \pi \hbar v_F T_\gamma] \not=0 $ for 
$ i \not= j $. This restriction implies  that the number of different phonons  
taking part in an electron-phonon scattering process is maximal.   
Eq.~(\ref{240}) delivers that for every phonon type of certain momentum 
the scattering probability is $ 1/2m_2 $ smaller than in the case of 
no existing chiral potential \cite{Dietel1}. By using (\ref{185}) we obtain 
for general step potentials within the same appromation used in 
(\ref{240}) $ \sum_{k_x} |\langle {\bf u}_s(x,q)|  \sigma e^{i k_x x}  
| {\bf u}_{s'}(x,q')\rangle|^2 \approx  1 $ for $ \sigma=\sigma_x $ and zero 
for $ \sigma \in \{\sigma_y, I\} $. This shows with the help  
of the discussion in
the second paragraph, that the considered 
chain of symmetric two-step potentials (\ref{240}) 
should give the best transport  
results over all step potentials and that even a general step potential 
should show an enhanced conductivity compared to the pristine NT.  
   
\begin{figure}
\vspace*{0.5cm} 
\begin{center}
\includegraphics[clip,height=8.0cm,width=8.0cm]{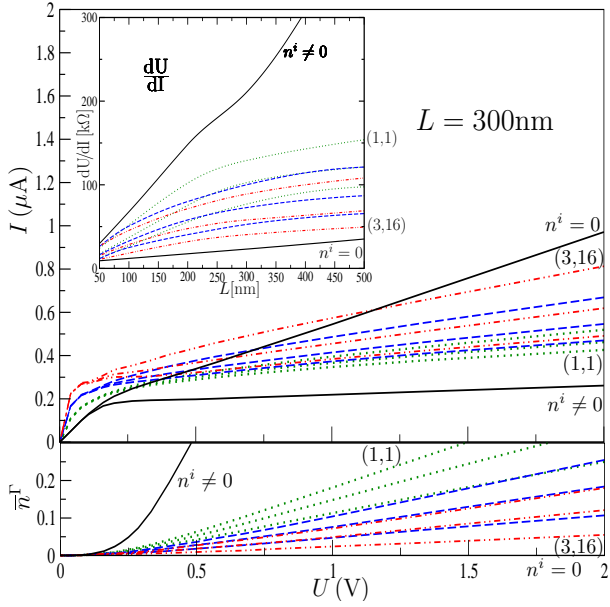}
\caption{(Color online) 
Upper panel: Current-voltage characteristic 
of the $(m_1,m_2) $ model ($2 m_1$ SL side-valleys and  
$ 2 m_2 $ phonon species) for $ m_1=1 $ (green dotted curves), 
$ m_1= 2 $ (blue dashed curves) and $ m_1= 3 $ (red dashed-dotted curves). 
Curves of the same style have different 
$ m_2 $ parameters with $ (m_1,1) $ (bottom curves), 
$ (m_1,m_1+1) $ (middle curves) and $ (m_1,(m_1+1)^2) $ (top curves).
This ordering is reversed in the inset and the lower panel. 
The NT length is $L=300 $nm.  
The black solid curves show the current-voltage 
characteristic with zero chiral potential $ V=0 $ and 
hot phonons ($n^i \not=0 $) or frozen phonons 
fixed at zero temperature ($n^i=0 $) \cite{Dietel1}. 
Inset: Differential resistance as a function of the NT length $ L $.
Lower panel:  
Position and energy averaged phonon number 
of SL inner-valley $ {\bf \Gamma} $-phonons 
$ \overline{n}^\Gamma $ for $ L=300$nm \cite{Dietel1}.      
}
\end{center}
\end{figure}

The number of SL valleys for a chain of two-step potentials is 
now $ m_1= [|t_\gamma| {\rm min}[V_i] d /T_\gamma \hbar v_F 2 \pi] $ 
which can be read off from Eqs.~(\ref{90}) and  
(\ref{100}) as in the case of the symmetric two-step 
potential.  Here we denote $ {\rm min} [V_i] $ 
as the minimum of all $ V_i $s.
Finally, we mention that for the forward scattering amplitudes,       
i.e. $ \partial_{q}\sin(\tilde{q}d/2)^2 \lessgtr 0 $ and 
$ \partial_{q'}\sin(\tilde{q}'d/2)^2 \lessgtr 0 $, we have 
$ (A^0_{\sigma_y})^2=(A^0_{I})^2=1 $, $ (A^0_{\sigma_x})^2=0 $ and 
$ (A^1_{\sigma})^2=0 $ in (\ref{240}).  

Until now, we have ignored transitions of electrons  
between the $ {\bf K} $ 
and $ {\bf K}' $ valleys. 
For the eigenvalue problem of the  
$ {\bf K}' $ valley we can repeat the discussion above for the 
$ {\bf K} $ valley by using the substitution 
$ t_\gamma \rightarrow -t_\gamma $ and $ q \rightarrow -q $ 
in the corresponding expressions \cite{Suzuura1}.  
Zone boundary $ A_1' $ phonon backward scattering is the only relevant 
phonon-scattering mechanism in this case \cite{Lazzeri1,Dietel1}.   
We now have to calculate the square of 
transition matrix element (\ref{240}) with   
$\sigma=\sigma_y$  \cite{Suzuura2} where one   
of the wavefunctions stands for a $ {\bf K}' $ valley 
function and the other is an eigenfunction 
of the $ {\bf K} $ valley. Eq.~(\ref{240}) leads to  
$ (A_{\sigma_y}^0)^2 \approx 0 $ and $ (A_{\sigma_y}^1)^2 \approx 1 $. The 
corresponding forward scattering amplitudes vanish.

Let us now consider the conductance of a NT with an SL potential. 
To this end we use the electron-phonon Boltzmann approach with 
the parameters established in 
Refs.~\onlinecite{Lazzeri1, Dietel1, Dietel2} for a NT lying on a substrate 
without chiral potential where also a thorough discussion on the scattering 
mechanisms can be found.  The leads and the substrate is  
fixed at room temperature. Before going into the details of our 
calculation we want to recall here 
that the number $ 2 m_1 $ was defined 
as the number of SL side-valley and $ 2m_2 $ are the number of 
steps in the SL chain potential which is equal to the number 
of different phonon species taking part in a backward scattering process.

In the following, we assume that higher bands do not contribute 
to the conductivity so that we have $ \hbar v_F T_\gamma 
 2 \pi/D   \gtrsim e U $ ($ \hbar v_F  
 2 \pi/d   \gtrsim e U $) for wavevectors of the SL out of (exact in) 
axial direction.   
For large enough applied bias voltages  
$  e U  \gg |\epsilon_s(q_i)| $ where $ q_i $ is defined by 
$ \sin(\tilde{q}_id/2 )=0 $, $ q^2_i \ll  T_\gamma^2 
({\rm min}[V_j] d /\hbar v_F)^2 $ 
and large chirality, i.e. $ t_\gamma  \gtrsim 1 $, 
we obtain the following idealized band system: 
We have one central SL band with dispersion $ \epsilon(k) \approx 
\pm \hbar v_F |k| $ and $ 2 m_1 $ SL sidebands with 
momentum shifted dispersion $ \epsilon(k) 
\approx \pm \lim_{k_0 \to 0} \hbar v_F \sqrt{k^2+ k^2_0} $. 
Phonons of $ 2m_2 $ types contribute to electron-phonon 
backward scattering with scattering times $ 2m_2 \tau_{\rm ep}^\nu $.
Here $ \nu $ stands for $ \Gamma $ 
for longitudinal $ E_2 $ zone-center scattering or $ K $ for 
$ A_1' $ zone boundary phonon scattering 
\cite{Dietel1, Lazzeri1}. 
$ \tau_{\rm ep}^\nu $ is the corresponding electron-phonon scattering time  
without chiral potential. We can simplify our calculation by 
using the same phonon velocity 
$ v^\nu_{\rm op} $ \cite{Dietel1, Lazzeri1} of the system without SL 
for all type of optical phonons. This is justified by the fact that our 
results do not depend much on the specific  
velocity value since the  
phonon mean free path is much smaller than the NT length as we  
have verified numerically.

It is enough to consider only forward scattering  
between the central SL valley and the $2 m_1 $ side valleys
mediated by transversal optical ${\bf \Gamma }$ phonons with 
scattering time $ \tau_{\rm ep}^\Gamma $ \cite{Dietel1} where we use calculation 
methods established in Ref.~\onlinecite{Dietel2} for forward scattering.  
At low voltages $ U\lesssim 0.17 V$, quasi-elastic scattering 
is relevant and we take it into account in our numerical calculations 
by inner SL valley scattering. This approximations is exact 
for voltages  lower than the SL side valley energy gap.     
We can simplify even further the model to an effective two-valley 
model with one central SL valley  and one side valley by using 
the approximation 
of periodic boundary conditions for the positions of the potential valleys 
in momentum space.

In Fig.~2, we show our results for the conductance, 
the differential conductance, 
and the position and energy averaged phonon number 
$ \overline{n}_\Gamma $, for certain 
$ (m_1,m_2)  $ values and lengths $ L $. Here 
$ n_\Gamma $ mediates 
the inner SL valley scattering.\mn{OK?}
Note that an upper limit for $2 m_1 $, $ 2 m_2 $ 
is given by the number of excitable circumverential phonons, 
i.e., $2 m_1, 2 m_2 \lesssim d/\sqrt{3} a$ where $ a $ is the NT 
interatomic distance $ a \approx 1.42 $\AA.
We obtain a strong increase in the absolute conductance and  
differential conductance at high voltages ($ U \approx 2 V $)
as a function of $ m_1 $ and $ m_2 $ while     
$ \overline{n}_\Gamma $ is strongly decreasing. 
The reason for an increase of the conductance for larger $ m_1 $ values 
and fixed $ m_2 $ comes mainly from the fact  that due to the 
band edges of the side valleys  scattering from the central SL 
valley to the SL side-valleys  is effectively forward. 
The backscattering to the 
central valley is accomplished then by a number of different phonons 
in contrast to the system without chiral potential.     
The growth of the conductance as a function of $m_1 $  
is then seen from our discussion in the second paragraph 
which also leads to the explanation of the 
conductance increase as a function of $ m_2 $. 

Summarizing, we have shown that large chiral 
unidirectional superlattice potentials 
in metallic NTs 
should lead to a large increase of the conductance, the differential
conductance,
and to a decrease of the optical phonon temperature at high voltage.
We have shown this explicly for a chain-like SL potential with 
symmetric steps. This kind of SL potential leads 
to the best transport results over all step potentials. Nevertheless,   
the main effect should be 
also observed for at least other SL step potentials with strong chirality.     
The effect arises from 
an increased number of phonons with 
different momenta but lower electron-phonon scattering 
probabilities contributing to  the electron-phonon scattering process.
As a result of our findings we expect an increase of the applicability 
of carbon NTs as
metallic wires.\mn{check}

\end{document}